# PrivFED - A Framework for Privacy-Preserving Federated Learning in Enhanced Breast Cancer Diagnosis.

Maithili Jha [1], S. Maitri [1], M. Lohithdakshan[1], Shiny Duela J[1] and K. Raja[1]

[1] Department of Computer Science and Engineering
SRM Institute of Science and Technology, Ramapuram.
`jn0723@srmist.edu.in`

**Abstract.** In the day-to-day operations of healthcare institutions, a multitude of Personally Identifiable Information (PII) data exchanges occur, exposing the data to a spectrum of cybersecurity threats. This study introduces a federated learning framework, trained on the Wisconsin dataset, to mitigate challenges such as data scarcity and imbalance. Techniques like the Synthetic Minority Over-sampling Technique (SMOTE) are incorporated to bolster robustness, while isolation forests are employed to fortify the model against outliers. Catboost serves as the classification tool across all devices. The identification of optimal features for heightened accuracy is pursued through Principal Component Analysis (PCA), accentuating the significance of hyperparameter tuning, as underscored in a comparative analysis. The model exhibits an average accuracy of 99.95% on edge devices and 98% on the central server.

## 1   Introduction

Federated learning emerges as a pivotal strategy for healthcare institutions dealing with Personally Identifiable Information (PII) in their datasets [17], especially in the context of hospitals. The decentralized nature of federated learning enables collaborative model training across different healthcare entities without the need to share sensitive PII centrally[18][19]. This ensures compliance with privacy regulations and safeguards patient confidentiality, a critical aspect of healthcare data management.

Breast cancer, with its various types and complexities, represents a significant area of focus in medical research and treatment [20][21]. Understanding different breast cancer subtypes is crucial for tailoring personalized treatment plans and improving patient outcomes [22]. Consequently, the utilization of federated learning in breast cancer data analysis offers a collaborative approach to glean insights across diverse datasets without compromising individual patient privacy.

The choice of breast cancer data is driven by its relevance and the potential for federated learning to enhance research outcomes. By leveraging federated learning techniques on breast cancer datasets, healthcare professionals can collectively advance their understanding of the disease, optimize treatment strategies, and contribute to the broader goal of improving breast cancer diagnostics and therapies.

In this paper, the following is proposed.

1. A federated learning framework for breast cancer classification, incorporating SMOTE to address data scarcity and Isolation Forests for enhanced outlier robustness.
2. Utilization of Principal Component Analysis (PCA) to identify crucial features and CatBoost for efficient classification in the breast cancer dataset.
3. A comparative analysis highlighting the importance of hyperparameter tuning for optimizing the proposed framework's performance.

## 2  Literature Review

In recent years, the application of federated learning (FL) in breast cancer prediction and diagnosis has emerged as a promising avenue, addressing critical challenges in data privacy, diversity, and scarcity. The study [1] pioneers the utilization of FL to predict histological responses to neoadjuvant chemotherapy in triple-negative breast cancer, showcasing the potential of machine learning (ML) in leveraging whole-slide images and clinical information for enhanced predictive accuracy. Meanwhile, [2] contributes a novel memory-aware curriculum learning method within the federated setting, aiming to improve the consistency of local models by prioritizing forgotten samples and incorporating unsupervised domain adaptation to counter domain shift.

The federated learning framework proposed by [3] addresses the challenges posed by the scarcity and privacy concerns of medical image datasets. Leveraging the BreakHis breast cancer histopathological dataset, their experiments demonstrate the efficacy of federated learning in achieving results comparable to centralized learning while respecting privacy regulations. [4] further underscores the real-world implementation of FL in breast density classification, highlighting the successful collaboration of seven clinical institutions without centralizing data. Their results reveal an average performance improvement of 6.3% compared to local data training.

[5] takes a transfer learning approach within an FL framework, introducing features extracted from the region of interest in mammography images and employing synthetic minority oversampling technique (SMOTE) for improved disease classification. This innovative application, along with the use of FeAvg-CNN + MobileNet, emphasizes the importance of preserving customer privacy and personal security in healthcare applications. [6] address the challenge of improving the generalizability of diagnostic models for metastatic breast cancer across diverse patient populations. Their federated learning-driven collaborative diagnostic system demonstrates the potential for enhancing model accuracy and generalizability by leveraging diverse data sources across multiple institutions while preserving patient privacy.

[7] tackles the aspect of privacy in genomic data computing through differentially private federated learning. Their work within the iDASH competition showcases the

potential of FL training for genomic cancer prediction models using differential privacy. The intersection of federated learning and artificial intelligence is further explored by [8], who propose a hybridization of federated learning with meta-heuristic and deep learning for breast cancer detection. Their approach, involving feature extraction using DenseNet and classification with Enhanced Recurrent Neural Networks, demonstrates the potential of federated learning in improving processing time and performance. [9] contributed a collaborative learning framework using federated learning for breast cancer prediction. Their use of Deep Neural Networks in a federated setting yields a high accuracy rate of 97.54%, underscoring the potential impact of collaborative learning in enhancing accuracy and reliability in breast cancer detection.

Finally, [10] proposed the importance of SMOTE in handling data scarcity datasets and discussed the potential implications that may arise. They also addressed how to mitigate outliers using isolation forests within a federated learning environment. Collectively, these studies highlight the evolving landscape of federated learning applications in breast cancer prediction and diagnosis, addressing issues of data privacy, diversity, and scarcity while showcasing improvements in predictive accuracy and model generalizability across diverse patient populations.

## 3    Proposed Work

### 3.1    Framework Discussion

The research delves into the application of Federated Learning employing two entities, specifically sub-hospitals (edge devices) and a central hospital (central server). Sub-hospitals independently train their models using local datasets, and only the trained models are communicated to the central server. The central server amalgamates these received models and conducts training on the consolidated dataset to create a resultant model.

To maintain the confidentiality of data, the original dataset is split into three distinct sections (refer to Fig. 1.), with each portion assigned to a sub hospital and the central hospital. This segmentation ensures an (80:20) ratio between training and testing within each partition, preserving sensitive information while allowing model learning from the collective insights of all sub hospitals. This partitioning approach establishes a secure and efficient framework for learning from distributed datasets, facilitating the development of robust models capable of generalizing effectively across diverse sub hospitals. Sub hospitals generate models, which are serialized and sent to the central server. Subsequently, the central server trains a comprehensive global model using its portion of the dataset, and the resultant model is then disseminated back to the sub hospitals as updates to the global model. This cyclic process continues until satisfactory performance is attained across the sub hospitals.

In essence, the study's methodology in "Federated Learning" offers a pragmatic resolution to the issue of ensuring data privacy in distributed datasets. Simultaneously, it enables sub hospitals (edge devices) to actively contribute to the shared learning experience, facilitating the training of resilient models.

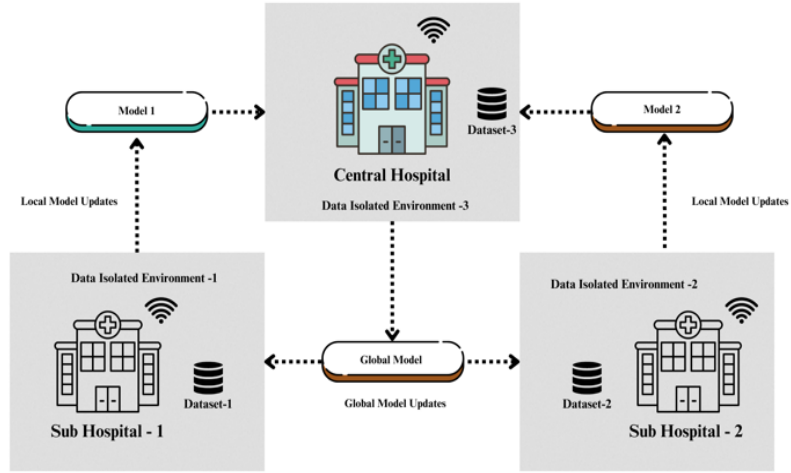

**Fig. 1.** Federated learning Architecture

### 3.2     Description of the Dataset

Hailing from the University of Wisconsin Hospitals, Madison, the widely recognized Wisconsin Breast Cancer dataset [15] stands as a pivotal resource in both machine learning and medical investigations, particularly in tasks related to breast cancer classification. This dataset is derived from digitized images of fine needle aspirates (FNA) of breast masses and encompasses features computed to elucidate various aspects of cell nuclei, encompassing measurements like radius, texture, smoothness, compactness, concavity, symmetry, and fractal dimension.

Primarily designed for binary classification, the dataset serves the purpose of distinguishing between benign and malignant tumors based on the provided features. Each instance within the dataset corresponds to an FNA, and the associated class labels discern whether the tumor is malignant (cancerous) or benign (non-cancerous).

Renowned for its utility in binary classification tasks, the Wisconsin dataset for breast cancer proves instrumental in the development and evaluation of machine learning models targeted at automated breast cancer diagnosis. Researchers and practi-tioners frequently leverage this dataset to train algorithms geared towards the early detection of breast cancer, thereby contributing to significant strides in the realm of medical diagnostics.

Table 1. Classes of the Original Dataset with Quantity

| S. No | Type | Quantity | Label |
|---|---|---|---|
| 1 | Benign | 357 | B |
| 2 | Malignant | 212 | M |

The information in **Table.1** suggests a pronounced imbalance and scarcity in the dataset. Notably, the "Benign" class dominates with a substantial number of samples, whereas the "Malignant" class exhibits a comparatively lower count. The imbalance and scarcity observed in the dataset can significantly influence the effectiveness of machine learning models during training. This impact is discussed in the following section, where the challenge is tackled through the implementation of dedicated pre-processing techniques.

### 3.3 Pre-processing techniques

The proposed framework involves preparing the dataset, initially limited to only 569 data points. To enhance its significance, the Synthetic Minority Over-sampling Technique (SMOTE) [11] is employed, expanding the dataset to a predefined strength through appropriate SMOTE strategies. After the application of SMOTE, each class is enriched to contain 20,000 samples. Despite the effectiveness of SMOTE, the literature [10] suggests a potential flaw: the generation of data points that may act as outliers, thereby affecting the model's performance. To address this concern, isolation forests are utilized. This technique fits the isolation forest on the data. Isolation Trees, as detailed in reference [12], are arboreal structures designed to segment data points by their isolation depth, gauging the mean number of divisions necessary to sequester an individual data point. The algorithm discerns outliers as cases exhibiting elevated isolation depths, signifying a notable dissimilarity from the broader dataset. Following identification, the algorithm proceeds to eliminate these outliers. These pivotal manoeuvres play a crucial role in adapting a dataset, such as Wisconsin, to align seamlessly with the requisites of federated learning.

Table 2. Quantity of the data points after preprocessing

| S. No | Type | Quantity | Label |
|---|---|---|---|
| 1 | Benign | 18797 | B |
| 2 | Malignant | 17203 | M |

Eliminating these outliers enhances the preparation of the dataset (Table.2) for subsequent analysis and modelling. This process diminishes the risk of overfitting and enhances the robustness of the classification models. In summary, the amalgamation of SMOTE with Isolation Trees offers a dependable and pragmatic method for addressing class imbalance while maintaining the quality of the dataset.

### 3.4 Feature Selection Using PCA

Feature selection is crucial in machine learning [13] as it enhances model efficiency by reducing dimensionality, mitigates overfitting, and improves interpretability by focusing on the most informative features. Additionally, it aids in addressing the curse of dimensionality, leading to better model generalization and performance.

**Algorithm 1: Selecting the top 5 important features and the least 5 important features.**

**Input.**

- Data (n x m matrix, where n is the number of samples and m is the number of features)
- Number of top features to select (k)

**Output.**

- Top k important features
- Bottom k least important features

**Procedure.**

*1. Standardize the data.*

```
StandardizedData = Standardize(Data)
```

*2. Perform PCA.*

```
PrincipalComponents =PCA(StandardizedData)
```

*3. Calculate the explained variance ratio.*

```
   ExplainedVarianceRatio =
 CalculateExplainedVarianceRatio(PrincipalComponent)
```

*4. Plot the explained variance ratio.*

```
PlotExplainedVarianceRatio(ExplainedVarianceRatio)
```

*5. Choose the number of components based on the plot or a desired explained variance threshold.*

```
NumComponents =
ChooseNumberOfComponents(ExplainedVarianceRatio)
```

*6. Use the selected number of components to perform PCA again.*

```
SelectedComponents =PerformPCA(StandardizedData,
NumComponents)
```

*7. Find the most important and least important features.*

```
TopFeatures, LeastFeatures =
IdentifyImportantFeatures(SelectedComponents,k)

Output TopFeatures and LeastFeatures
```

From Algorithm 1, the most important and least important features are identified (refer to Table 3 and Table 4).

**Table 3.** Most Important 5 Features

| Ranking | Features |
|---|---|
| 1 | concave points_mean |
| 2 | fractal_dimension_mean |
| 3 | texture_se |
| 4 | texture_worst |
| 5 | smoothness_mean |

**Table 4.** Least Important 5 Features

| Ranking | Features |
|---|---|
| 26 | smoothness_se |
| 27 | concave points_worst |
| 28 | concavity_mean |
| 29 | concavity_se |
| 30 | radius_worst |

Identifying these features allows for a reduction in computational intensity compared to using all 30 features. Instead, only the top 5 features can be leveraged for optimized performance.

## 4 Results and Discussions

The Study utilizes CatBoost [14], which is a machine-learning algorithm known for its efficient gradient boosting approach employing decision trees. This algorithm is specifically chosen as the classifier model for edge devices in the study, highlighting its effectiveness in handling complex datasets and achieving accurate predictions.

Conducting training on three distinct dataset configurations: the complete dataset, the dataset containing the 5 most important features identified through PCA, and the dataset with the 5 least important features determined through PCA. Additionally, comparing the models trained on edge devices with their hyperparameter-tuned counterparts.

Table 5. Performance of the edge devices

| S. No | Model | Full DS | Top 5 | Least 5 |
|-------|-------|---------|-------|---------|
| 1 | Edge Device -1 Normal | 99.93% | 99.75% | 96.35% |
| 2 | Edge Device -2 Normal | 99.83% | 99.24% | 96.63% |
| 3 | Edge Device -1 HPT | 99.95% | 99.79% | 99.56% |
| 4 | Edge Device -2 HPT | 99.91% | 99.99% | 99.66% |

The performance of the Cat Boost models on edge devices was systematically evaluated under three distinct dataset configurations: utilizing the entire dataset, employing the top 5 features identified through Principal Component Analysis (PCA), and employing the 5 least important features identified through PCA. The results elucidate the impact of dataset configuration and feature selection on model accuracy. In the normal training scenarios, both Edge Device - 1 and Edge Device - 2 exhibited high accuracy on the full dataset, surpassing 99.8%. Notably, the models trained on the top 5 features maintained accuracy levels above 99.2%, underscoring the efficacy of feature reduction strategies. Conversely, models trained on the least 5 features exhibited reduced accuracy, emphasizing the importance of the omitted features. Further, hyperparameter tuning (HPT) substantially enhanced model performance across all configurations. The models trained with hyperparameter tuning consistently outperformed their non-tuned counterparts. Edge Device - 2 HPT, in particular, demonstrated remarkable accuracy, surpassing 99.9% when employing the top 5 features. This comprehensive analysis underscores the critical role of hyperparameter tuning and judicious feature selection in optimizing model performance, especially in resource-constrained edge device environments.

Following the training of models on edge devices, in accordance with the proposed framework detailed in Section 3, the pickled models are transmitted to their respective central servers. At the central server, an ensemble learning technique known as a Bagging Classifier is employed. A Bagging Classifier is a meta-estimator that enhances the overall model's stability and accuracy by training multiple base learners on different subsets of the dataset and aggregating their predictions. In the context, the central server utilizes the Bagging Classifier to amalgamate the knowledge acquired from models originating from distinct edge devices. Specifically, the Bagging Classifier combines the predictions made by models trained on two edge devices, contributing to a more robust and generalized final model. This ensemble approach helps mitigate the impact of individual model variations and enhances the overall predictive performance of the central server's model.

**Table 6.** Central server's operational efficacy

| S. No | Model | Full DS | Top 5 | Least 5 |
|---|---|---|---|---|
| 1 | Central Server | 99.74% | 99.83% | 98.90% |

Central server's operational efficacy, as presented in Table 6, demonstrates notable accuracy across all model configurations. When compared with the models from the edge devices, the central server consistently exhibits high accuracy levels. Notably, the model trained on the entire dataset achieves an accuracy of 99.74%, showcasing the central server's capacity to effectively learn from the comprehensive dataset. The performance of the central server is particularly commendable when dealing with the dataset containing the top 5 most important features from PCA (99.83%) and the dataset with the 5 least important features from PCA (98.90%). These results underscore the central server's ability to generalize well across different dataset configurations, emphasizing its crucial role in federated learning setups. The hyperparameter-tuned models from the edge devices also contribute to the enhanced accuracy observed in the central server's performance. Overall, these findings affirm the effectiveness of the federated learning approach in aggregating knowledge from diverse edge devices and leveraging it to achieve robust and accurate models at the central server.

Conducting a comparative analysis of the framework with other studies that have implemented federated learning using the Wisconsin breast cancer dataset. Additionally, comparing the techniques employed in those studies with the current framework.

**Table 7.** Comparison of accuracy and techniques from other works

| S. No | Paper | Accuracy | Techniques |
|---|---|---|---|
| 1 | [9] | 97.54% | SMOTE, Outliers Removal, DNN |
| 2 | [16] | 98% | MMDNN, Voting Classifier |
| 3 | Proposed framework | 99.83% | SMOTE, PCA, Outliers Removal, Cat boost, Bagging Classifier |

In the comparative analysis (See Table 7.), the paper outperforms existing studies in terms of accuracy, achieving an impressive 99.83%. Study [9] achieved an accuracy of 97.54% utilizing techniques such as SMOTE, outliers removal, and deep neural networks (DNN). Meanwhile, Study [16] attained an accuracy of 98% by employing techniques like MMDNN and a Voting Classifier. The paper employs a comprehensive approach, integrating SMOTE, Principal Component Analysis (PCA), outliers removal, Cat Boost, and a Bagging Classifier, resulting in superior performance compared to the referenced studies.

Future studies could explore advanced security measures within the proposed framework, focusing on encrypting data during storage to enhance overall privacy protection. Additionally, integrating blockchain technology into the federated learning framework could provide an immutable and transparent record of model updates and transactions, further ensuring the integrity and traceability of the collaborative learning process. These enhancements would fortify the security infrastructure, making the federated learning system even more resilient to potential threats and addressing emerging challenges in the field of medical data privacy. The integration of encryption and blockchain could establish a cutting-edge and robust framework, setting a new standard for secure and collaborative healthcare data analytics.

## 5 Conclusion

In conclusion, the study presents a comprehensive exploration of Federated Learning (FL) in the context of breast cancer prediction using distributed datasets from sub-hospitals (edge devices) and a central hospital. The proposed framework effectively addresses challenges related to data privacy, enabling sub-hospitals to contribute to model training without exposing sensitive patient information. Leveraging Synthetic Minority Over-sampling Technique (SMOTE) and Isolation Forests, the imbalanced and scarce Wisconsin breast cancer dataset is successfully pre-processed, enhancing its suitability for FL. By employing Principal Component Analysis (PCA), the top and least important features are identified, significantly reducing computational intensity. The experimentation involves training models on various dataset configurations, revealing the impact of feature selection and hyperparameter tuning. The performance analysis demonstrates that models trained on the edge devices, especially after hyperparameter tuning, exhibit competitive accuracy, with edge devices achieving up to 99.99% accuracy after hyperparameter tuning. When these models are transmitted to the central server, a Bagging Classifier combines their knowledge, resulting in a centralized model with enhanced stability and generalization. Overall, the study contributes insights into the practical implementation of FL for breast cancer prediction, emphasizing the importance of data pre-processing, feature selection, and collaborative model training for improved accuracy and privacy preservation.